\documentclass[reprint,amsmath,amssymb,aps,prl,twocolumn,showpacs,superscriptaddress]{revtex4-1}

\usepackage{graphicx}	
\begin{document}

\title{Ultrafast Optical Excitation of a Persistent Surface-State Population in the Topological Insulator Bi$_{2}$Se$_{3}$}

\author{J.~A. Sobota}
\author{S. Yang}
\affiliation{Stanford Institute for Materials and Energy Sciences, SLAC National Accelerator Laboratory, 2575 Sand Hill Road, Menlo Park, CA 94025, USA}
\affiliation{Geballe Laboratory for Advanced Materials, Department of Applied Physics, Stanford University, Stanford, CA 94305, USA}
\affiliation{Department of Physics, Stanford University, Stanford, CA 94305, USA}
\author{J.~G. Analytis}
\affiliation{Stanford Institute for Materials and Energy Sciences, SLAC National Accelerator Laboratory, 2575 Sand Hill Road, Menlo Park, CA 94025, USA}
\affiliation{Geballe Laboratory for Advanced Materials, Department of Applied Physics, Stanford University, Stanford, CA 94305, USA}
\author{Y.~L. Chen}
\affiliation{Stanford Institute for Materials and Energy Sciences, SLAC National Accelerator Laboratory, 2575 Sand Hill Road, Menlo Park, CA 94025, USA}
\affiliation{Geballe Laboratory for Advanced Materials, Department of Applied Physics, Stanford University, Stanford, CA 94305, USA}
\affiliation{Department of Physics, Stanford University, Stanford, CA 94305, USA}
\author{I.~R. Fisher}
\affiliation{Stanford Institute for Materials and Energy Sciences, SLAC National Accelerator Laboratory, 2575 Sand Hill Road, Menlo Park, CA 94025, USA}
\affiliation{Geballe Laboratory for Advanced Materials, Department of Applied Physics, Stanford University, Stanford, CA 94305, USA}
\author{P.~S. Kirchmann}
\email{kirchmann@fhi-berlin.mpg.de}
\affiliation{Stanford Institute for Materials and Energy Sciences, SLAC National Accelerator Laboratory, 2575 Sand Hill Road, Menlo Park, CA 94025, USA}
\affiliation{Fritz Haber Institute of the Max Planck Society, Department of Physical Chemistry, Faradayweg 4-6, D-14195 Berlin, Germany}
\author{Z.-X. Shen}
\email{zxshen@stanford.edu}
\affiliation{Stanford Institute for Materials and Energy Sciences, SLAC National Accelerator Laboratory, 2575 Sand Hill Road, Menlo Park, CA 94025, USA}
\affiliation{Geballe Laboratory for Advanced Materials, Department of Applied Physics, Stanford University, Stanford, CA 94305, USA}
\affiliation{Department of Physics, Stanford University, Stanford, CA 94305, USA}


\begin{abstract}

Using femtosecond time- and angle- resolved photoemission spectroscopy, we investigated the nonequilibrium dynamics of the topological insulator Bi$_2$Se$_3$.  We studied $p$-type Bi$_2$Se$_3$, in which the metallic Dirac surface state and bulk conduction bands are unoccupied.  Optical excitation leads to a meta-stable population at the bulk conduction band edge, which feeds a nonequilibrium population of the surface state persisting for $>$10~ps.  This unusually long-lived population of a metallic Dirac surface state with spin texture may present a channel in which to drive transient spin-polarized currents.

\end{abstract}
\pacs{78.47.J-, 73.20.-r, 72.25.Fe, 79.60.-i}
\maketitle


Three dimensional topological insulators (TI) are fascinating compounds that combine an insulating bulk electronic band structure with a surface state (SS) that crosses the Fermi level $E_F$ \cite{Zhang2009}. The SS has novel properties such as topological protection \cite{Fu2007,Qi2008}, suppression of backscattering \cite{Roushan2009}, and a helical spin texture \cite{Hsieh2009a,Hsieh2009b}. Its response to optical excitation is expected to lead to several interesting phenomena, such as large Kerr rotation \cite{Qi2008,Aguilar2011} and surface spin current \cite{Hosur2011}.  However, few experimental studies have focused on the SS response due to the fact that surface signatures are easily overwhelmed by bulk contributions to transport \cite{Analytis2010a,Butch2010} and optical measurements \cite{Qi2010,Kumar2011}.  

\begin{figure}
\resizebox{\columnwidth}{!}{\includegraphics{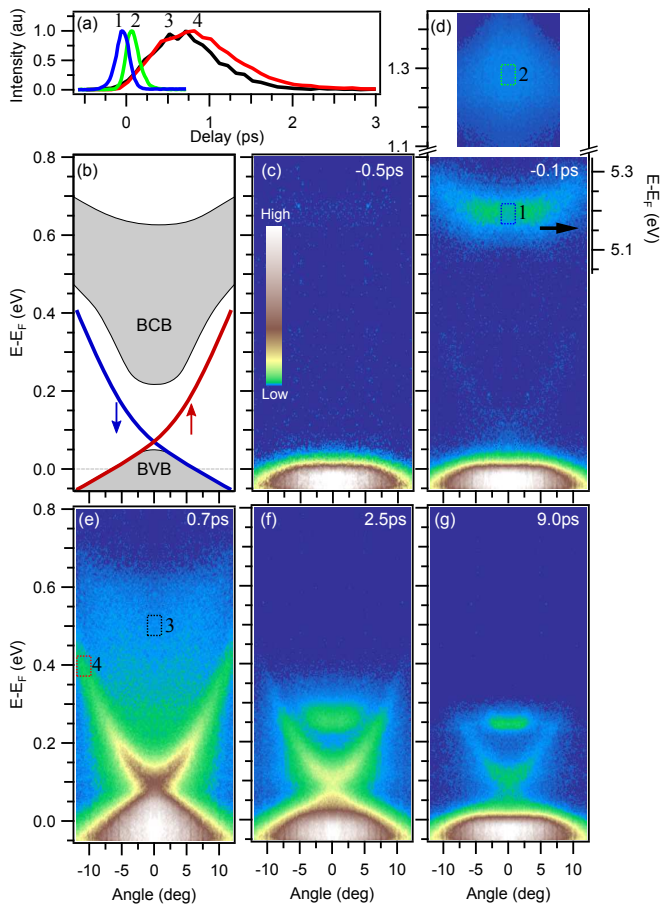}}
\caption{(Color online) (a) Transient photoemission intensity within the integration windows indicated in the subsequent panels.  Feature (1) corresponds to an IPS populated by $h\nu_2$ and probed by $h\nu_1$ and thus appears before time-zero.  Feature (2) is a bulk state populated by a direct optical transition at time-zero.  The BCB (3) and SS (4) are indirectly populated by scattering from higher energy states.  (b) Schematic of the electronic band structure for Bi$_2$Se$_3$. Grey, shaded regions represent the BVB and BCB. Lines represent the SS with spin-texture indicated. (c)-(g) trARPES spectra near the $\Gamma$-point for various pump-probe delays. (c) BVB before excitation. (d) Initial optical transition to high-lying bulk state and IPS. Note the separate energy axis for the IPS. (e) The SS and BCB are populated by scattering from higher-lying states. (f) Energy relaxation of the SS and BCB populations are mostly complete. (g) A meta-stable population in the BCB, accompanied by a persistent population of the SS in the region energetically below it.  A movie showing data at all delays is included in the supplemental material \cite{SOM}.
\label{fig1}}
\end{figure}

In this Letter we report time- and angle- resolved photoemission spectroscopy (trARPES) experiments on the topological insulator Bi$_2$Se$_3$ and elucidate the electron dynamics in response to optical excitation \cite{Haight1995,Sch08,Roh11}. The femtosecond time, energy, and momentum resolution allows us to probe fundamental scattering processes directly within the electronic band structure, and thus distinguish the separate dynamics of the bulk and surface. We use $p$-doped samples to start with a completely unoccupied SS.  Optical excitation populates high-lying bulk states, which rapidly decay to lower energy states via inter- and intra-band phonon-mediated scattering processes.  Within 2~ps a meta-stable population forms at the bulk conduction band (BCB) edge.  This population acts as an electron reservoir which fills the SS with a steady supply of carriers for $\sim$10~ps.  The persistent occupation of a metallic state is a unique situation, since the absence of a bandgap in metals typically means that there is no barrier to rapid recombination.  Further, the SS has the same Dirac dispersion seen in equilibrium experiments, where a novel spin-momentum locking has been demonstrated \cite{Hsieh2009b}.  We propose that this long-lived spin-textured metallic population presents a channel in which to drive transient spin-polarized currents.


Our trARPES setup is described in detail in the supplemental material \cite{SOM}. Single crystals of Bi$_2$Se$_3$ were synthesized by slow cooling a binary melt, with nominal Bi:Se ratio of 35:65. $p$-type Bi$_2$Se$_3$ was achieved by adding small amounts of Mg to the melt ($<$0.5\% of the Bi molar amount). The carrier density of these samples was measured from the Hall and Shubnikov-de Haas effect to be 2.3$\times$10$^{18}$ cm$^{-3}$.  The samples are cleaved \emph{in-situ} under a base pressure $<1\times10^{-10}$ torr. We optically excite the sample with a linearly-polarized infrared laser pulse ($h\nu_1=1.5$~eV, $50$~fs) and subsequently probe the evolution of the transient populations in the occupied and unoccupied band structure by photoemitting electrons with an ultraviolet laser pulse ($h\nu_2=6.0$~eV, $160$~fs). The photoelectron kinetic energy and emission angle are resolved by a hemispherical electron analyzer. During measurement the sample temperature is maintained at 70~K.  Our laser setup is based on a Ti:Sapphire oscillator operating at $80$~MHz repetition rate. In this particular experiment, we compromise the ultraviolet pulse duration for a good spectral resolution of $<22$~meV.  The time resolution obtained by cross-correlation of $h\nu_1$ and $h\nu_2$ is 163~fs.  The absorbed infrared fluence is computed to be 26~$\mu$J$/$cm$^2$ using the measured reflectivity of 0.4 at 45$^{\circ}$ incidence \cite{SOM}.


Fig.~\ref{fig1}(a) displays the transient photoemission intensity within the integration windows indicated in the subsequent panels.  A movie of the same data is included in the supplemental material \cite{SOM}.  For reference, a schematic of the band structure is illustrated in Fig.~\ref{fig1}(b).  At negative delays --- before the arrival of the infrared pump pulse --- we probe the equilibrium electronic structure. Due to the $p$-type doping, the SS and BCB are unoccupied in equilibrium, see Fig.~\ref{fig1}(c). The first features to be populated are displayed in Fig.~\ref{fig1}(d).  Curiously, the parabolic dispersing feature (labeled 1) appears to be populated before time-zero.  Moreover, closer inspection reveals that it decays toward negative delays \cite{SOM}.  We attribute this feature to the first image potential state (IPS) \cite{Fau95,Petek1998,Haight1995,Wolf1996,Echenique1978}, which is an electron state bound in front of the metallic Bi$_2$Se$_3$ surface, and frequently observed in time-resolved photoemission experiments.  Its decay toward negative delays indicates that it is populated by $h\nu_2$ and probed by $h\nu_1$, hence its separate energy axis in the figure.  We compute a binding energy $E_\text{IPS} -E_\text{vac} = -0.77(3)$~eV, consistent with the IPS binding energy $-0.85$~eV expected for a perfectly metallic surface \cite{SOM,Fau95,Petek1998}.  Importantly, the decay toward negative delay implies that the IPS relaxation is decoupled from the SS and bulk band dynamics which are the focus of this Letter, and we can safely disregard it for the remaining discussion.

The feature at $E-E_F=1.3$~eV (labeled 2) decays with a time constant $70(20)$~fs and results from a direct optical transition pumped by $h\nu_1$ from the BVB edge to higher lying states in the bulk.  Note that the lower-lying BCB and SS have negligible population around time-zero, and it takes $\sim700$~fs for them to reach maximum population, see Fig.~\ref{fig1}(a).  This indicates that these bands are not directly populated by $h\nu_1$, but are populated indirectly by scattering from higher-lying states. The SS dispersion agrees with ARPES measurements in thermal equilibrium, see Fig.~\ref{fig1}(e) \cite{Hsieh2009b,Chen2010}, ensuring that we can discuss transient populations in a fixed electronic structure. The BCB does not have a sharp dispersion, consistent with its 3D bulk nature, but appears as a diffuse electron distribution centered around the $\Gamma$-point. 

After $~2$~ps the SS and BCB populations have significantly decayed and energetically relaxed towards the bottom of their respective bands, see Fig.~\ref{fig1}(f). The subsequent dynamics is much slower and persists for $>10$~ps.  Obstructed from further decay due to the bandgap, the BCB electrons form a meta-stable population at the BCB edge.  Intriguingly, this BCB population is accompanied by a persistent population in the SS, but only energetically below the BCB edge, see Fig.~\ref{fig1}(g).

\begin{figure}
\resizebox{\columnwidth}{!}{\includegraphics{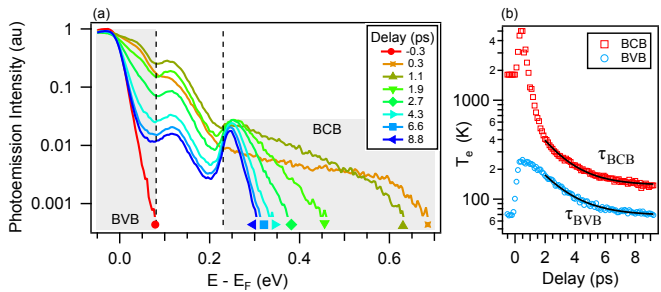}}
\caption{(Color online) (a) EDC at the $\Gamma$-point for various delays.  The energetic regions associated with the BVB and BCB are specified.  (b) The transient electronic temperatures extracted by fitting Fermi-Dirac distributions to the EDCs.  After 2~ps the BVB and BCB temperatures relax exponentially with  time constants $\tau_\text{BVB} = 1.85(6)$~ps and $\tau_\text{BCB}= 1.67(5)$~ps, respectively.
\label{fig2}}
\end{figure}

To understand the coupled dynamics of bulk and surface electrons in more detail, we proceed by analyzing the energetic distribution of electrons in the bulk. Fig.~\ref{fig2}(a) presents energy distribution curves (EDCs), obtained by integrating the trARPES spectral intensity within a  $\pm2$$^{\circ}$ angular window around the $\Gamma$-point, for selected delays.  We first focus on the energetic region of the EDCs associated with the BVB.  By fitting each EDC with a Fermi-Dirac (FD) distribution, we extract an electronic temperature $T_e$ as a function of delay, shown in Fig.~\ref{fig2}(b) \cite{SOM}.  At negative delays there is a well-defined FD distribution at $E_F$ in the BVB. After excitation, a FD with increased $T_e$ is observed and attributed to scattering of photoexcited electrons with electrons in the cold Fermi sea \cite{Giu05,Qui58}.  The hot electron population subsequently cools by transferring energy to the lattice \cite{Ani74,Bovensiepen2007}.  Within 9~ps, $T_e$ almost returns to its equilibrium value, and the BVB electrons and lattice system can be regarded as equilibrated.  We find that $T_e$ decays exponentially with a time constant $\tau_\text{BVB} = 1.85(6)$~ps.  Note that we confine our fit to delays after 2~ps to avoid non-analytic transients associated with indirect filling processes.

We now turn to the energetic region of the EDCs associated with the BCB.  The BCB is unpopulated before excitation, but within $300$~fs a diffuse electron distribution extends $>0.5$~eV above $E_F$.  The electrons are distributed exponentially in energy, suggesting that intra-band scattering processes rapidly (within our 160~fs time resolution) redistribute the electrons within the band, taking them into a thermal distribution.  The distribution subsequently cools, again by transferring energy to the lattice.  By fitting to a FD distribution we characterize the BCB population with an effective temperature, shown in Fig.~\ref{fig2}(b) \cite{SOM}.  Fitting $T_e$ after 2~ps, as above, we find an exponential decay with time constant $\tau_\text{BCB}= 1.67(5)$~ps.  The near agreement of $\tau_\text{BVB}$ and $\tau_\text{BCB}$ motivates us to identify their mean $\tau_\text{e-p} = 1.74(4)$~ps as the timescale for electrons to transfer energy to the lattice, and is thus a characteristic time for electron-phonon scattering in the bulk of Bi$_2$Se$_3$.  We note that similar timescales for intra-band cooling have been observed in optical pump-probe measurements \cite{Qi2010, Kumar2011, Hsieh2011a}.

\begin{figure}
\resizebox{\columnwidth}{!}{\includegraphics{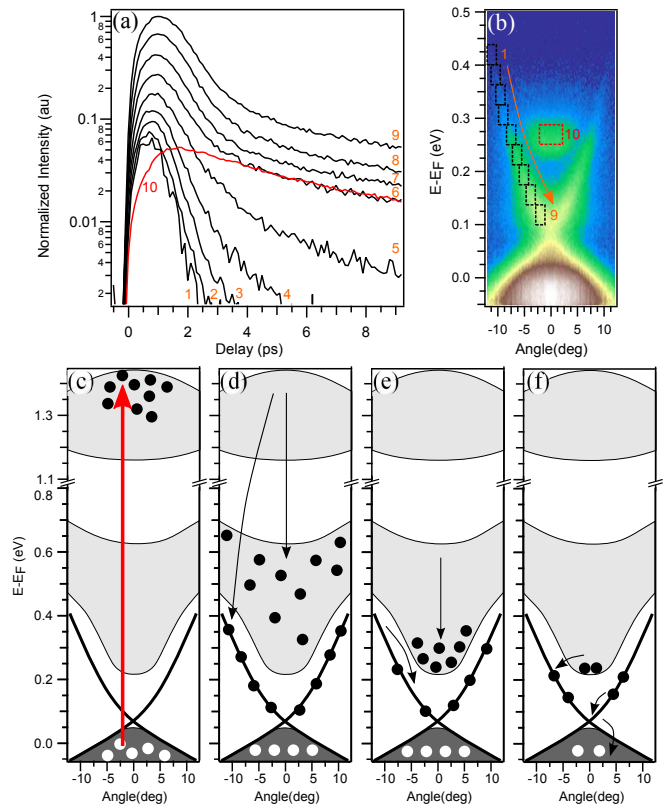}}
\caption{(Color online) (a) Transient photoemission intensities within the integration windows indicated in the subsequent panel (b).  Curve \#10 corresponds to the integration window over the BCB edge, and is normalized to match the intensity of the SS intensities.  The SS population decay is exponential at energies above the BCB edge, but has a second, slower component below the BCB edge.  This slow component decays with the same timescale of 5.95(2)~ps as the BCB population, demonstrating that the SS is continuously filled by the slowly decaying BCB.  (c-f)  Schematic of the transitions and scattering processes discussed in this Letter, including the (c) direct optical transition, (d) scattering into SS and BCB, (e) intra-band scattering of the BCB and SS, and (f) the BCB-to-SS scattering responsible for the persistent SS population. 
\label{fig3}}
\end{figure}
After this energy-relaxation process, the system is characterized by a meta-stable BCB population.  This long-lived BCB population is not unusual, and is in fact responsible for the phenomenon of photoconductivity which is typically observed in semiconductors \cite{Sze2006}.  The longevity of this population is due to the fact that the electron-phonon scattering processes responsible for energy relaxation within the band are incapable of recombining electrons with the BVB.  Here the highest phonon energy is 23~meV \cite{Richter1977}, while the bandgap of 200~meV is an order of magnitude larger \cite{Chen2010}.  This slow recombination of bulk carriers in Bi$_2$Se$_3$ has previously been observed in optical measurements \cite{Hsieh2011a}.  The simultaneous persistence of the metallic SS population, however, is surprising.  Photoconductivity of a metallic state is unexpected since the absence of a bandgap means there is no barrier to rapid recombination. We attribute this SS population to a continuous filling from the meta-stable BCB population.  The first evidence for this is Fig.~\ref{fig1}(g), in which it is evident that the SS persists only energetically below the BCB edge.  We  find yet stronger evidence in the population dynamics.  In Fig.~\ref{fig3}(a) we plot the transient photoemission intensity obtained by integrating within the windows indicated in Fig.~\ref{fig3}(b).  Above the BCB, the SS decays with a single exponential.  However, below the BCB a second slower component is observed, reflecting the presence of an additional filling channel.  Moreover, the decay rate of this slower component matches that of the population decay at the BCB edge (see curves \#6 and \#10).  The agreement of these decay rates is direct evidence that the persistent SS population is due to continuous filling from the meta-stable BCB population.  We note that similar bulk-to-surface scattering behavior has been observed in Si, which however lacks a metallic SS crossing $E_F$ \cite{Weinelt2004,Tanaka2009}.

We can deduce some properties of the transiently excited populations relevant to transport applications.   For bulk crystals in equilibrium, current is typically dominated by non-spin-polarized BCB carriers rather than the spin-polarized SS \cite{Analytis2010a,Butch2010}.  In our experiment, the ratio of photoexcited SS carriers to BCB carriers is enhanced with respect to the equilibrium case due to the finite optical penetration depth of the excitation ($\alpha^{-1} = 50$~nm) \cite{SOM}.  Since the transient populations are generated only in this near-surface region, their transport properties are more comparable to those of thin films and nanoribbons, in which SS contributions are indeed observed \cite{Steinberg2010,Peng2010}.  Thus, the photoexcited SS population could represent a channel through which to drive transient spin currents without overwhelming contributions from the bulk.

An exponential fit to the BCB population decay (curve \#10 in Fig.~\ref{fig3}(a)) gives a population lifetime of $5.95(2)$~ps.  Since the SS population is sustained by the BCB, it is relevant to investigate the mechanism for the BCB decay. We have argued that electron-phonon coupling cannot be responsible for recombination due to the low phonon energies.  Direct recombination via photon emission is also a possibility, though the timescale for such a process is typically $\gg 1$~ns, and thus seems unlikely to be relevant on our measurement timescale \cite{Schroder2005}.  The BCB decay could also be attributed to spatial diffusion of BCB electrons away from the probed volume of the sample.  The diffusion coefficient $D$ for Bi$_2$Se$_3$ has been measured to be 1.2~cm$^2$/s \cite{Kumar2011}.  Combining this with the absorption coefficient of the pump $\alpha = 0.02$~nm$^{-1}$ \cite{SOM}, we estimate a characteristic diffusion time $(\alpha^2D)^{-1} = 21$~ps \cite{Rowe1993}.  This is rapid enough for diffusion to be a significant contribution to our measured lifetime.  Characterization and control of spatial diffusion, for example by utilizing thin films with thickness less than the pump penetration depth, could therefore drastically extend the observed lifetime.

In conclusion, we have performed trARPES measurements on $p$-type Bi$_2$Se$_3$ and elucidated the rich electron dynamics generated by optical excitation.  We have identified the first image potential state in front of the metallic surface, which may be of particular interest in this material due to the proposal that TIs host image magnetic monopoles \cite{Qi2009}.  We have also identified the direct transitions and subsequent relaxation processes induced by optical excitation, which we have schematically illustrated in Fig.~\ref{fig3}(c)-(f).  Finally, we have shown that these transitions culminate in a persistent non-equilibrium population of the spin-textured Dirac SS, which is attributed to continuous filling from a meta-stable population in the BCB.  We note that excitation with any above-bandgap photon energy should lead to the same SS filling behavior, since inter- and intra-band scattering processes inevitably lead to relaxation of carriers toward the BCB edge.  This phenomenon could find a role in applications requiring ultrafast optical control of a spin-polarized surface conduction channel.

\begin{acknowledgments}
The authors thank R.~G. Moore, F. Schmitt, W.~S. Lee, Z.~K. Liu, and T.~P. Devereaux for helpful discussions.  J.~A.~S. acknowledges support by the Stanford Graduate Fellowship.  P.~S.~K. acknowledges support by the Alexander-von-Humboldt foundation through a Feodor-Lynen fellowship and continuous support by M. Wolf.  This work is supported by the Department of Energy, Office of Basic Energy Sciences, Division of Materials Science.
\end{acknowledgments}

\bibliography{BibTex}

\end{document}